\documentclass{article}

\usepackage[final]{neurips_2024}

\usepackage[utf8]{inputenc}
\usepackage[T1]{fontenc}
\usepackage[hidelinks]{hyperref}
\usepackage{url}
\usepackage{booktabs}
\usepackage{amsfonts}
\usepackage{amsmath}
\usepackage{amssymb}
\usepackage{amsthm}
\usepackage{nicefrac}
\usepackage{microtype}
\usepackage{xcolor}
\usepackage{graphicx}
\usepackage{subcaption}
\usepackage{algorithm}
\usepackage{algorithmic}
\usepackage{multirow}
\usepackage{tikz}
\usepackage{pgfplots}
\usepackage{bbm}
\usepackage{listings}
\usepackage{enumitem}
\usepackage{float}
\pgfplotsset{compat=1.18}
\usepgfplotslibrary{fillbetween}
\usetikzlibrary{patterns,positioning,shapes.geometric,arrows.meta,calc,fit,backgrounds}

\newcommand{\taac}{\textsc{TAAC}}
\newcommand{\plexor}{\textsc{Plexor}}

\newcommand{\ind}{\mathbbm{1}}

\newtheorem{finding}{Finding}
\newtheorem{hypothesis}{Hypothesis}
\newtheorem{proposition}{Proposition}

\renewcommand{\And}{%
  \end{tabular}\hfil\linebreak[0]\hfil%
  \begin{tabular}[t]{c}%
}

\title{Evaluating Small Language Models for Front-Door Routing:\\A Harmonized Benchmark and Synthetic-Traffic Experiment}

\author{
  Warren Johnson\thanks{Corresponding author. E-mail: \texttt{warrenjo@plexor.dev}}\\
  Plexor Labs\\
  Sammamish, WA, USA
  \And
  Charles Lee\\
  Project Autobots\\
  Seattle, WA, USA
}

\date{}

\begin{document}

\maketitle

\begin{abstract}
Selecting the appropriate model at inference time---the routing problem---requires jointly optimizing output quality, cost, latency, and governance constraints. Existing approaches delegate this decision to LLM-based classifiers or preference-trained routers that are themselves costly and high-latency, reducing a multi-objective optimization to single-dimensional quality prediction. We argue that small language models (SLMs, 1--4B parameters) have now achieved sufficient reasoning capability for sub-second, zero-marginal-cost, self-hosted task classification, potentially making the routing decision negligible in the inference budget. We test this thesis on a six-label taxonomy through two studies. Study~1 is a harmonized offline benchmark of Phi-3.5-mini, Qwen2.5-1.5B, and Qwen-2.5-3B on identical Azure T4 hardware, serving stack, quantization, and a fixed 60-case corpus. Qwen-2.5-3B achieves the best exact-match accuracy (0.783), the strongest latency--accuracy tradeoff, and the only nonzero accuracy on all six task families. Study~2 is a pre-registered four-arm randomized experiment under synthetic traffic with an effective sample size of 60 unique cases per arm, comparing Phi-4-mini, Qwen-2.5-3B, and DeepSeek-V3 against a no-routing control. DeepSeek-V3 attains the highest accuracy (0.830) but fails the pre-registered P95 latency gate (2,295\,ms); Qwen-2.5-3B is Pareto-dominant among self-hosted models ($0.793$ accuracy, 988\,ms median, \$0 marginal cost). No model meets the standalone viability criterion ($\geq 0.85$ accuracy, $\leq 2{,}000$\,ms P95). The cost and latency prerequisites for SLM-based routing are met; the accuracy gap of 6--8 percentage points and the untested question of whether correct classification translates to downstream output quality bound the remaining distance to production viability.
\end{abstract}

\vspace{0.3em}

\vspace{0.5em}
\noindent\rule{\textwidth}{0.4pt}
\vspace{0.3em}
\noindent\textbf{Series position.} This is Article~8 in the \taac{} Research Series. Articles 1--3 established task-dependent compression thresholds and the perplexity paradox mechanism \citep{johnson2026compress,johnson2026perplexity,johnson2026cliff}. Articles 4--5 documented provider-dependent output token explosion under compression \citep{johnson2026greenai,johnson2026benchmark}. Article~6 validated benchmark-derived thresholds in a production RCT ($N = 358$, six arms, CONSORT reporting) \citep{johnson2026rct}. Article~7 designed tokenizer-aware prompt languages achieving up to 64.7\% token reduction \citep{johnson2026promptlang}. This article shifts from \emph{what to compress} to \emph{which model should decide}---extending the ``route'' branch of the Article~1 dichotomy into model selection. The current study replaces the earlier pilot benchmark and generational bridge studies with a harmonized three-model benchmark eliminating cross-study confounds and reports results of a four-arm randomized experiment under synthetic traffic.
\vspace{0.5em}
\noindent\rule{\textwidth}{0.4pt}

% ════════════════════════════════════════════════════════════════════
% 1. INTRODUCTION
% ════════════════════════════════════════════════════════════════════
\newpage
\section{Introduction}
\label{sec:intro}

Selecting the appropriate model for a given request at inference time has become a central challenge in production LLM deployment \citep{jiang2023routellm,shnitzer2023large,chen2023frugalgpt}. The economic motivation is substantial: at 100,000 queries per day, routing simple requests to cheaper models while reserving frontier capacity for complex tasks can yield annualized savings exceeding \$150,000 \citep{lu2023routing}. But routing is itself an inference problem, and current solutions are expensive and incomplete. Learned routers train on preference data to predict which model produces better output \citep{jiang2023routellm,ong2025routellm}, but the training signal is single-dimensional---output quality---and the router's own inference adds latency and cost. Cascade architectures invoke multiple models per request, compounding latency \citep{chen2023frugalgpt,madaan2023automix,yue2024cascades}. Production frameworks commonly delegate routing to another LLM, simplifying the problem to categorization---but this approach is itself costly, does not address speed or governance, and ignores the operational reality that models competitive on general-purpose benchmarks often diverge sharply on task-specific quality \citep{ding2024hybrid,shnitzer2023large}. More fundamentally, these approaches reduce routing to a single optimization axis---``which model gives better output?''---without jointly addressing the multiple priorities that production routing must balance: output quality, inference cost, response latency, compute capacity, and data governance.

We argue that small language models (SLMs) in the 1--4B parameter range have now crossed a capability threshold that transforms the economics of this decision. A self-hosted SLM running on a commodity GPU can classify prompt intent in sub-second latency at zero marginal per-request cost\footnote{``Zero marginal cost'' denotes the absence of variable per-request API billing, not zero total cost of ownership. Self-hosted SLMs incur fixed infrastructure cost (${\sim}\$0.75$/hr GPU rental) that must be amortized; the per-request cost advantage over commercial APIs is realized at scale.}, with no external API dependency and full data residency. If these models are accurate enough, the front-door routing decision becomes negligible in the inference cost and latency budget---freeing the system to optimize across all routing dimensions simultaneously rather than trading router cost against routing quality. The question is whether current SLMs have advanced enough to realize this shift: whether they can reason over task signals with sufficient accuracy to serve as production routers. This paper provides the first controlled evaluation.

The front-door is itself a model-selection problem. SLMs offer sub-second latency but can fail at nuanced task-taxonomy boundaries; stronger SLMs or even full-scale models improve classification quality but may violate latency budgets, introduce external API dependencies, or negate cost savings by consuming the very inference budget they are meant to protect \citep{chen2023frugalgpt}. This tradeoff is amplified when teams require both \emph{scientific validity}---paired significance tests, dataset immutability, reproducibility---and \emph{operational validity}---health checks, logging schema, alertability, arm isolation. The tension between these two modes of validity is rarely acknowledged in the routing literature, where most work evaluates routing quality in isolation from the operational infrastructure that must sustain it in production \citep{madaan2023automix}.

The challenge is further compounded by the rapid pace of SLM development. The transformer architecture \citep{vaswani2017attention} has been adapted into increasingly efficient small-scale variants, from the Phi family's textbook-quality training data approach \citep{gunasekar2023phi1,li2023phi15,javaheripi2023phi2,abdin2024phi3,abdin2024phi4} to the Qwen family's multilingual scaling strategy \citep{yang2024qwen2,yang2024qwen25} and the open-weight ecosystem of Llama \citep{touvron2023llama2,meta2024llama3}, Gemma \citep{gemma2024gemma2,gemma2024gemma}, Mistral \citep{jiang2023mistral,jiang2024mixtral}, TinyLlama \citep{zhang2024tinyllama}, StableLM \citep{zhu2024stablelm}, and the TinyStories line of research demonstrating surprisingly coherent generation from sub-billion-parameter models \citep{eldan2023tinystories}. Each new release reshuffles the Pareto frontier of quality versus cost, rendering previously optimal routing configurations suboptimal within months. This ``model churn'' problem means that any static evaluation of routing models is inherently time-limited---a finding that motivates our multi-study design, which progresses from a locked offline benchmark through a generational bridge experiment to a pre-registered randomized experiment that can accommodate future model candidates.

Article~1 in this series \citep{johnson2026compress} demonstrated a fundamental dichotomy: code-generation tasks benefit from compression-first strategies (Cohen's $d = 1.84$ over 2,650 trials), while chain-of-thought reasoning benefits from routing to appropriately-capable models. That finding established \emph{routing} as a first-class cost-reduction mechanism parallel to compression. But it left open the question of \emph{which model performs the routing itself}---the subject of the present article. The distinction is important: while prior work on learned routing \citep{jiang2023routellm,ong2025routellm} trains a dedicated classifier on preference data, and cascade architectures \citep{madaan2023automix,yue2024cascades} invoke multiple models per query, our front-door approach requires a single SLM to make an interpretable, discrete classification decision using only the raw prompt text. This constraint---single model, single pass, discrete output---places unique demands on SLM capability that differ from the generative and reasoning tasks on which SLMs are typically benchmarked \citep{hendrycks2021mmlu,clark2018arc,zellers2019hellaswag,bigcode2023evaluation,austin2021mbpp}. The question of which SLM should serve as the front-door classifier also intersects with emerging work on test-time compute scaling \citep{snell2024testtime} and speculative decoding \citep{leviathan2023speculative,chen2023speculative}, suggesting that the traditional tradeoff between model size and inference speed may be more nuanced than a simple parameter-count comparison implies.

We contribute two complementary studies spanning the full path from offline evaluation through production deployment: (1) a \textbf{harmonized offline benchmark} (Study~1, \S\ref{sec:methods1}--\S\ref{sec:results1}) comparing three SLMs on identical hardware, serving stack, and corpus with pairwise McNemar tests; and (2) an \textbf{executed four-arm randomized experiment under synthetic traffic} (Study~2, \S\ref{sec:rct-design}--\S\ref{sec:rct-results}) with $N = 400$ per arm ($n_{\text{eff}} = 60$ for accuracy; see \S\ref{sec:rct-results}), SHA-256 session-ID randomization, and a pre-registered decision matrix with viable-region gates. Beyond the empirical findings, this paper makes three methodological contributions: a harmonized evaluation protocol that eliminates cross-study hardware, framework, and corpus confounds (\S\ref{sec:methods1}), drawing on best practices from clinical trial design \citep{schulz2010consort} and trustworthy online experimentation \citep{kohavi2020trustworthy,kohavi2009experiments}; a pre-registered multi-arm randomized experiment adapted from online A/B testing methodology \citep{kohavi2020trustworthy}, including Bonferroni correction for six hypotheses, O'Brien-Fleming interim boundaries \citep{obrien1979}, and sequential testing principles \citep{wald1945}; and a transparent reporting of negative results---no arm meets the pre-registered viable region---with actionable deployment recommendations derived mechanically from the decision matrix.

\subsection{The Model Selection Problem for Front-Door Routing}
\label{sec:model-selection}

The model routing problem has emerged as a central challenge in production LLM deployment, and we identify four dominant architectural paradigms. \emph{Learned routers} train on preference data to predict which model will produce a better response for a given prompt. \citet{jiang2023routellm} achieve $2\times$ cost reduction on MT-Bench while maintaining 95\% of GPT-4 quality with a BERT-scale classifier \citep{devlin2019bert}; \citet{ong2025routellm} extend this to an open-source framework; and \citet{hu2024routerbench} provide a comprehensive benchmark for evaluating such multi-LLM routing systems. \emph{Benchmark-guided routing} \citep{shnitzer2023large} selects the model with the highest predicted performance on a prompt's task type using benchmark scores as a prior, building on holistic evaluation frameworks \citep{liang2023helm,srivastava2023beyond}. \emph{Cascade architectures} \citep{chen2023frugalgpt,madaan2023automix,yue2024cascades,lu2023routing} invoke cheaper models first and escalate to expensive models only when confidence is low, differing from our single-pass front-door approach. \emph{Meta-modeling approaches} \citep{sakota2024flyswat,zhang2024onthefly} predict cost-effectiveness of each candidate for a given input; these are conceptually closest to our work but operate at the model-selection level rather than the task-classification level. Speculative decoding \citep{leviathan2023speculative,chen2023speculative}, Branch-Train-Merge \citep{li2022branchtrain}, and Mixture-of-Agents \citep{wang2024moa} address inference efficiency through parallel execution rather than routing. Most recently, \citet{ding2024hybrid} frame routing explicitly as a cost--quality tradeoff, training a quality predictor to decide between a smaller and larger model per query; their approach optimizes along two dimensions (cost and quality) but requires a trained predictor per model pair and does not address latency, governance, or multi-model taxonomies with more than two routing targets.

Our work addresses a specific variant that we term \emph{front-door classification}: a single SLM classifies each incoming prompt into a discrete task taxonomy, and the taxonomy label deterministically selects the downstream model and compression policy. This differs from learned routing in three ways. First, the classification label is human-readable, auditable, and overridable---learned routers produce opaque preference scores. Second, the taxonomy label can trigger multiple downstream decisions (model tier, compression ratio, output budget, quality gate) through a lookup table, whereas learned routers optimize for a single downstream metric. Third, the taxonomy is version-controlled and can be updated without retraining. The tradeoff is that front-door classification requires a well-defined taxonomy with clear boundaries; our 6-family taxonomy (\texttt{code/simple}, \texttt{code/complex}, \texttt{CoT/simple}, \texttt{CoT/complex}, \texttt{hybrid/agentic}, \texttt{hybrid/generative}) is derived from the Article~1 dichotomy extended with complexity and modality dimensions.

The use of SLMs in the 1--4B parameter range as task classifiers is motivated by several factors: classification requires understanding prompt \emph{intent} rather than generating a complete response, potentially requiring less parametric capacity \citep{kaplan2020scaling}; the output budget is minimal (a single JSON label), eliminating the output token dynamics that dominate inference cost for generative tasks \citep{willard2023outlines,zheng2024outlines}; and SLMs can be self-hosted on commodity GPUs, avoiding API dependencies and data-residency concerns. However, SLMs face known limitations. The Phi family \citep{abdin2024phi3,abdin2024phi4} achieves strong performance on structured tasks but may struggle with ambiguous prompts. The Qwen family \citep{yang2024qwen2,yang2024qwen25} exhibits strong multilingual capabilities but has received less evaluation on classification-specific tasks. DeepSeek's MoE architecture \citep{liu2024deepseek,dai2024deepseek} activates only 37B of its 671B parameters per token, but the internal routing decision may interact with the classification task unpredictably \citep{shazeer2017moe,fedus2022switch,lepikhin2021gshard}. Recent surveys of small language models \citep{lu2024slmsurvey,wang2024slmsurvey} provide comprehensive coverage but do not specifically address classification-for-routing. The Orca \citep{mukherjee2023orca,mitra2023orca2} and WizardLM \citep{xu2023wizardlm} lines of work demonstrate that smaller models can acquire sophisticated reasoning through progressive learning from stronger models, and parameter-efficient adaptation methods such as prefix tuning \citep{li2021prefix} and prompt tuning \citep{lester2021prompt} offer low-cost specialization paths. Production deployment of SLM classifiers requires 4-bit quantization to fit within GPU memory budgets on commodity hardware (T4, A10); recent work on QLoRA \citep{dettmers2023qlora}, GPTQ \citep{frantar2023gptq}, AWQ \citep{lin2024awq}, and SmoothQuant \citep{xiao2023smoothquant} has demonstrated that 4-bit quantization preserves most quality for generative tasks, but the impact on classification accuracy specifically is less well-characterized.

A natural question is whether generative SLMs are the right tool for classification-for-routing at all. Fine-tuned discriminative models---BERT \citep{devlin2019bert}, DeBERTa, or distilled variants---trained on even modest labeled datasets ($n \geq 200$) typically achieve high classification accuracy with sub-100\,ms latency and minimal compute. \citet{jiang2023routellm} use a BERT-scale classifier to achieve $2\times$ cost reduction, suggesting that discriminative fine-tuning is a viable and potentially superior approach. We deliberately evaluate zero-shot generative SLMs in this paper because (1)~the taxonomy evolves with the product, and retraining a discriminative classifier for each taxonomy change incurs maintenance cost; (2)~generative SLMs can output structured JSON with confidence signals, enabling richer routing logic; and (3)~the zero-shot setting provides a lower bound on SLM capability that fine-tuning can only improve. However, we acknowledge that this paper does not include a fine-tuned discriminative baseline, which limits our ability to assess whether the generative SLM approach offers advantages beyond flexibility. A direct comparison between zero-shot Qwen-2.5-3B and a DeBERTa-base model fine-tuned on the 60-case corpus (via cross-validation) is planned as a critical ablation for the next article in this series.

\subsection{Beyond Classification: The Multi-Objective Routing Gap}
\label{sec:multi-objective}

Existing routing approaches share a common structural limitation: they reduce routing to classification along a single axis---typically output quality or task type---without jointly optimizing across the multiple objectives that production routing must satisfy. In practice, the routing decision must balance at least five dimensions: (1)~\emph{output quality}---does the selected model produce responses that meet the task's quality requirements, given that surface-level benchmark scores are unreliable proxies for task-specific performance? (2)~\emph{inference cost}---what is the total cost of the routed request, including both the front-door classification and the downstream model, accounting for costly retries when a cheaper model fails quality requirements? (3)~\emph{response latency}---does the total pipeline latency (front-door $+$ downstream) meet the application's SLA, and does the router itself contribute minimal overhead? (4)~\emph{capacity and availability}---is the selected model available and able to absorb the routed traffic? (5)~\emph{governance and data residency}---can the request be processed without sending data to external APIs that may violate compliance constraints?

The SLM-as-classifier architecture addresses dimensions 2--5 structurally: self-hosted SLMs have zero marginal cost, sub-second latency, deterministic capacity, and full data residency. The open question is dimension~1---not merely whether the SLM classifies \emph{accurately} (the subject of the present paper), but whether accurate classification translates to \emph{downstream output quality}. A router that correctly labels a prompt as \texttt{code/complex} but routes it to a model that produces inadequate code has achieved label accuracy without routing value. Just because a downstream model is cheaper does not mean it produces the level of quality required---a point that applies equally to the routing decision itself. This distinction between classification accuracy and routing effectiveness motivates a two-phase evaluation design. The present paper addresses \textbf{Phase~1}: can SLMs classify accurately and quickly enough to serve as candidate routers? This is a necessary condition. \textbf{Phase~2}---testing whether correct classification translates to downstream output quality and cost savings---requires executing the full routing pipeline and is deferred to future work (\S\ref{sec:future}). Establishing Phase~1 first is methodologically sound: if SLMs cannot classify with sufficient accuracy, the routing thesis fails regardless of downstream quality.

\subsection{Research Questions and Hypotheses}
\label{sec:hypotheses}

The overarching research question is: \emph{Which small language model backend maximizes front-door classification accuracy while satisfying latency and cost constraints for production routing?} We decompose this into hypotheses organized by study.

\textbf{Study~1 hypotheses (validated).}

\begin{hypothesis}[H1: Quality Ordering]
\label{hyp:quality}
\emph{Under the harmonized evaluation contract (same hardware, same serving stack, same corpus), the three SLMs produce significantly different exact-label accuracy, with pairwise McNemar tests identifying which pairs differ.}
\end{hypothesis}

\begin{hypothesis}[H2: Latency-Quality Tradeoff]
\label{hyp:latency-s1}
\emph{The three SLMs exhibit distinct latency profiles, with smaller models achieving lower latency but potentially lower accuracy.}
\end{hypothesis}

\begin{hypothesis}[H3: Task-Family Interaction]
\label{hyp:family}
\emph{The quality advantage is not uniform across task families: different SLMs dominate different families, producing a ``split policy'' rather than a single winner.}
\end{hypothesis}

\begin{hypothesis}[H4: Family Coverage]
\label{hyp:coverage}
\emph{The 3B Qwen model achieves non-zero accuracy on all six task families, demonstrating breadth of coverage that smaller models lack.}
\end{hypothesis}

\textbf{Study~2 hypotheses (pre-registered; partially testable under synthetic traffic).}

\begin{hypothesis}[H5: Routing Accuracy (Treatment Arms Only)]
\label{hyp:routing}
\emph{Routing accuracy differs across treatment arms B, C, and D, with expected ordering $D > C \geq B$. Arm~A (control) has undefined routing accuracy and is excluded from this endpoint; comparisons of treatment arms against Arm~A test the end-to-end impact of routing on cost and quality under separate endpoints (H6, H8). Success threshold: $\geq 85\%$ for the winning treatment arm. Pairwise tests: three comparisons (B\,vs.\,C, B\,vs.\,D, C\,vs.\,D) with Holm-Bonferroni correction.}
\end{hypothesis}

\begin{hypothesis}[H6: Total Cost]
\label{hyp:cost}
\emph{Average USD per request (front-door $+$ downstream) differs, with $B$ or $C < D < A$. Threshold: $\geq 25\%$ reduction vs.\ Arm~A.}
\end{hypothesis}

\begin{hypothesis}[H7: Classification F1]
\label{hyp:f1}
\emph{F1 macro and hybrid/agentic-specific F1 differ, with $D > C \geq B$. Threshold: F1 macro $\geq 0.82$.}
\end{hypothesis}

\begin{hypothesis}[H8: Latency SLA]
\label{hyp:latency-rct}
\emph{P95 front-door latency differs, with $C \leq B < D$ (network penalty). Hard gate: $\leq 2{,}000$\,ms for all arms.}
\end{hypothesis}

\begin{hypothesis}[H9: Compression Quality]
\label{hyp:compression}
\emph{Average compression ratio at semantic similarity $\geq 0.88$ differs by model backend. Threshold: $\geq 30\%$ token reduction.}
\end{hypothesis}

% ════════════════════════════════════════════════════════════════════
% 2. METHODS: STUDY 1 (PILOT BENCHMARK)
% ════════════════════════════════════════════════════════════════════
\section{Methods: Study~1 (Harmonized Offline Benchmark)}
\label{sec:methods1}

Earlier pilot studies reported in preliminary versions of this work used separate hardware, frameworks, GPU allocation modes, and corpus SHAs, introducing cross-study confounds. The harmonized benchmark eliminates all confounds by running all three models on the same hardware, same stack, same corpus, in the same session. This section describes the experimental contract---the complete specification that must be held invariant for results to be valid---the models under evaluation, the classification system prompt, and the statistical analysis plan.

\textbf{Formal problem statement.}\quad We define the front-door classification task as follows. Let $\mathcal{P}$ denote the space of natural-language prompts and let the label set be
\[
  \mathcal{L} = \{\texttt{cd/smp},\;\texttt{cd/cplx},\;\texttt{CoT/smp},\;\texttt{CoT/cplx},\;\texttt{hyb/agnt},\;\texttt{hyb/gen}\}.
\]
A front-door classifier is a function
\begin{equation}
  f: \mathcal{P} \longrightarrow \mathcal{L} \times [0,1],
  \label{eq:classifier}
\end{equation}
mapping each prompt $p$ to a predicted label $\hat{\ell} = f(p)_{\text{label}} \in \mathcal{L}$ and a self-reported confidence $c = f(p)_{\text{conf}} \in [0,1]$. The downstream routing table $R: \mathcal{L} \to \mathcal{M} \times \Theta$ maps each label to a (model, compression-policy) pair, making the classification decision the single point of control for the entire inference pipeline.

\textbf{Experimental contract.}\quad The contract identifier is \texttt{jsonsafe\_cls\_tokens\_128}. The corpus is \texttt{progressive\_test\_\allowbreak{}cases\_v2\_60.jsonl} (60 prompts, 6 families $\times$ 10 cases each; SHA-256: \texttt{dac3aac5\ldots}). Decoding: \texttt{max\_new\_tokens}=128, $T = 0.0$ (greedy). Quantization is 4-bit NF4 via bitsandbytes for all three models. Fairness invariants require sequential execution with exclusive GPU allocation, the same serving stack, and the same corpus SHA across all runs.

The three models span the SLM design space: Phi-3.5-mini-instruct (3.8B, dense, Microsoft) optimizes for instruction-following quality; Qwen2.5-1.5B-Instruct (1.5B, dense, Alibaba) optimizes for minimal footprint; Qwen-2.5-3B-Instruct (3.0B, dense, Alibaba) \citep{yang2024qwen25} tests whether a generation jump within the Qwen family resolves the accuracy deficit.

\textbf{Classification system prompt.}\quad All three models receive an identical, frozen classification system prompt:

\begin{quote}
\small
\begin{verbatim}
You are a task classifier for an AI routing system.
Classify the prompt into exactly one of these categories:
- code/simple (single function, snippet, trivial script)
- code/complex (multi-file, architecture-level, tests required)
- CoT/simple (single-step explanation or reasoning)
- CoT/complex (multi-step reasoning, trade-off analysis)
- hybrid/agentic (autonomous execution, self-healing,
  multi-artifact, no-confirmation)
- hybrid/generative (mixed creative + structured output)

Key rule: autonomous mode + no confirmation + multi-artifact = hybrid/agentic.

Respond with JSON only: {"label": "<label>", "confidence": <0.0-1.0>}
\end{verbatim}
\end{quote}

This prompt was frozen as of \texttt{benchmark\_openweights.py} line 347 and used identically across both studies. The decision to use a zero-shot prompt without chain-of-thought reasoning \citep{wei2022chain,kojima2022zeroshotreasoners} was deliberate: chain-of-thought prompting for classification would increase output token count and latency, potentially negating the front-door's cost advantage. Few-shot prompting \citep{brown2020language,min2022incontext} was similarly excluded to maintain prompt brevity and cross-model comparability.

\textbf{Inference framework.}\quad All three models were served by vLLM 0.17.1 \citep{kwon2023vllm} with bitsandbytes 4-bit NF4 quantization backend. Using a single serving stack for all models eliminates inference-framework confounds.

\textbf{Execution environment.}\quad Microsoft Azure Standard\_NC8as\_T4\_v3, NVIDIA Tesla T4 (16\,GB GDDR6). All three models were executed sequentially with exclusive GPU allocation (no concurrent workloads). Corpus SHA-256: \texttt{dac3aac5\ldots}. Latencies are measured end-to-end from prompt submission to JSON response receipt on the vLLM serving stack, including tokenization and output decoding but excluding production-path overhead (load balancing, authentication, logging) that would add latency in deployment.

\textbf{Statistical analysis plan.}\quad The primary endpoint is exact label correctness per case (paired, binary), tested with pairwise exact McNemar two-sided \citep{mcnemar1947} for all three model pairs (Phi vs.\ Qwen-1.5B, Phi vs.\ Qwen-3B, Qwen-1.5B vs.\ Qwen-3B). This is the strictest possible metric---a partially correct label (e.g., correct task type but wrong complexity level) receives zero credit. Secondary endpoints are parse success rate and per-case latency (median and P95). Confidence intervals use the bootstrap percentile method with 10,000 resamples \citep{efron1979bootstrap}. Family-level analyses are exploratory (10 cases per family) and are reported without multiple-comparison correction.

% ════════════════════════════════════════════════════════════════════
% 3. RESULTS: STUDY 1
% ════════════════════════════════════════════════════════════════════
\section{Results: Study~1 (Harmonized Benchmark)}
\label{sec:results1}

Given a labeled evaluation set $\mathcal{D} = \{(p_i, \ell_i)\}_{i=1}^{N}$, we measure exact-match accuracy as
\begin{equation}
  \text{Acc}(f, \mathcal{D}) \;=\; \frac{1}{|\mathcal{D}|} \sum_{(p,\ell)\in\mathcal{D}} \ind\bigl[f(p)_{\text{label}} = \ell\bigr],
  \label{eq:accuracy}
\end{equation}
where $\ind[\cdot]$ is the indicator function. This is the strictest possible metric: partial matches (e.g., correct task type but wrong complexity) receive zero credit.

Table~\ref{tab:pilot-overall} presents the harmonized 60-case benchmark results. All three models were run sequentially on the same Azure Standard\_NC8as\_T4\_v3 GPU with exclusive allocation, identical vLLM 0.17.1 serving stack, and identical bitsandbytes 4-bit NF4 quantization. Qwen-2.5-3B achieves the highest accuracy ($0.783$, $47/60$) with perfect parse rate and $1{,}088$\,ms median latency. No significant difference was detected between Phi-3.5-mini and Qwen-3B in accuracy (exact McNemar $p = 0.503$, 20 discordant pairs: 8 favoring Phi, 12 favoring Qwen-3B), but Phi-3.5-mini is $5.3\times$ slower (median $5{,}772$\,ms vs.\ $1{,}088$\,ms). Both significantly outperform Qwen2.5-1.5B ($p = 0.0009$ and $p < 0.0001$ respectively). The harmonized accuracies for Phi-3.5-mini ($0.717$) and Qwen-1.5B ($0.400$) are substantially higher than earlier pilot results ($0.517$ and $0.200$ respectively on RunPod A40 with HuggingFace \texttt{model.generate()}), which we attribute to the shift from unoptimized HuggingFace inference to vLLM serving with PagedAttention---suggesting that inference framework choice materially affects classification output even at temperature $T = 0.0$.

\begin{table}[htbp]
\centering
\caption{Harmonized 60-case benchmark metrics (Azure T4, vLLM 0.17.1, bitsandbytes 4-bit NF4, exclusive GPU, 2026-03-15/16).\protect\footnotemark}
\label{tab:pilot-overall}
\begin{tabular}{lccc}
\toprule
Metric & Phi-3.5-mini & Qwen2.5-1.5B & Qwen-2.5-3B \\
\midrule
Exact classification accuracy & $0.7167^{**}$ & 0.4000 & $\mathbf{0.783}^{***}$ \\
JSON parse rate & 0.9833 & 0.9667 & $\mathbf{1.0000}$ \\
Median latency (ms) & 5,772 & $\mathbf{793}$ & 1,088 \\
P95 latency (ms) & 6,357 & 4,636 & $\mathbf{1{,}554}$ \\
\bottomrule
\end{tabular}
\end{table}
\footnotetext{Significance convention used throughout all tables: $^{*}$ $p < 0.05$; $^{**}$ $p < 0.01$; $^{***}$ $p < 0.001$. Bold indicates best per metric. Pairwise McNemar tests: Phi vs.\ Qwen-1.5B $p = 0.0009$; Phi vs.\ Qwen-3B $p = 0.503$ (ns); Qwen-1.5B vs.\ Qwen-3B $p < 0.0001$.}

\begin{finding}[Quality-Latency Tradeoff]
\label{find:tradeoff}
\emph{At $n = 60$, no significant accuracy difference was detected between Qwen-2.5-3B and Phi-3.5-mini ($p = 0.503$); however, statistical power is limited (${\approx}50\%$ for a 15\,pp MDE), so this reflects insufficient evidence to distinguish the models rather than confirmed equivalence. Qwen-2.5-3B is $5.3\times$ faster (median $1{,}088$\,ms vs.\ $5{,}772$\,ms), making it the Pareto-dominant model on the cost-of-error criterion. Neither meets the 85\% production threshold.}
\end{finding}

\textbf{Family-level behavior.}\quad The family-level results in Table~\ref{tab:pilot-family} reveal complementary specialization patterns. Phi-3.5-mini achieves perfect accuracy on \texttt{code/complex} (1.00) and \texttt{CoT/complex} (1.00) but scores only 0.50 on \texttt{hybrid/agentic} and 0.20 on \texttt{hybrid/generative}. Qwen2.5-1.5B achieves perfect accuracy on \texttt{hybrid/agentic} (1.00) but fails on all CoT families (0.00). Qwen-2.5-3B is the only model achieving non-zero accuracy on all six families, with its strongest performance on \texttt{CoT/complex} (0.90), \texttt{CoT/simple} (0.90), and \texttt{hybrid/agentic} (0.90). Its weakest family is \texttt{hybrid/generative} (0.60), which remains the most challenging family for all models.

\begin{table}[htbp]
\centering
\caption{Family-level exact-match accuracy (10 cases per family), harmonized benchmark.}
\label{tab:pilot-family}
\begin{tabular}{lcccc}
\toprule
Family & Phi-3.5-mini & Qwen2.5-1.5B & Qwen-2.5-3B & Best Model \\
\midrule
\texttt{code/complex} & $\mathbf{1.00}$ & 0.50 & 0.70 & Phi \\
\texttt{code/simple} & 0.80 & $\mathbf{0.90}$ & 0.70 & Qwen-1.5B \\
\texttt{CoT/complex} & $\mathbf{1.00}$ & 0.00 & 0.90 & Phi \\
\texttt{CoT/simple} & 0.80 & 0.00 & $\mathbf{0.90}$ & Qwen-3B \\
\texttt{hybrid/agentic} & 0.50 & $\mathbf{1.00}$ & 0.90 & Qwen-1.5B \\
\texttt{hybrid/generative} & 0.20 & 0.00 & $\mathbf{0.60}$ & Qwen-3B \\
\bottomrule
\end{tabular}
\end{table}

% ---- Figure 1: Accuracy-Latency Scatterplot ----
\begin{figure}[t]
\centering
\begin{tikzpicture}
\begin{axis}[
    width=0.85\textwidth,
    height=0.55\textwidth,
    xlabel={P95 Latency (ms)},
    ylabel={Exact Classification Accuracy},
    xmin=0, xmax=7000,
    ymin=0, ymax=1.05,
    xtick={0,1000,2000,3000,4000,5000,6000,7000},
    xticklabel style={/pgf/number format/fixed, /pgf/number format/1000 sep={,}},
    ytick={0,0.2,0.4,0.6,0.8,1.0},
    legend pos=south east,
    grid=none,
    clip=false,
]
% Viable region (shaded green)
\fill[green!15] (axis cs:0,0.85) rectangle (axis cs:2000,1.05);
\node[anchor=south west, font=\scriptsize\bfseries, green!50!black] at (axis cs:50,0.86) {Viable};

% Production threshold line
\draw[dashed, red!70!black, thick] (axis cs:0,0.85) -- (axis cs:7000,0.85);
\node[anchor=south west, font=\scriptsize, red!70!black] at (axis cs:4500,0.855) {Accuracy = 0.85};

% Latency SLA line
\draw[dashed, blue!70!black, thick] (axis cs:2000,0) -- (axis cs:2000,1.05);
\node[anchor=south west, font=\scriptsize, blue!70!black, rotate=90] at (axis cs:2050,0.10) {P95 $\leq$ 2{,}000\,ms};

% Study 1 (Harmonized Benchmark) — filled markers, P95 values
\addplot[only marks, mark=*, mark size=4pt, red!80!black] coordinates {(4636,0.400)};
\node[anchor=north west, font=\scriptsize] at (axis cs:4700,0.39) {Qwen-1.5B};

\addplot[only marks, mark=triangle*, mark size=5pt, teal!80!black] coordinates {(1554,0.783)};
\node[anchor=south west, font=\scriptsize] at (axis cs:1620,0.785) {Qwen-3B};

\addplot[only marks, mark=square*, mark size=4pt, blue!80!black] coordinates {(6357,0.717)};
\node[anchor=south west, font=\scriptsize] at (axis cs:6420,0.72) {Phi-3.5};

% Study 2 (Experiment) — open markers, P95 values
\addplot[only marks, mark=square, mark size=4pt, blue!80!black] coordinates {(1541,0.518)};
\node[anchor=north east, font=\scriptsize] at (axis cs:1480,0.50) {Phi-4$^{\text{E}}$};

\addplot[only marks, mark=triangle, mark size=5pt, teal!80!black] coordinates {(1170,0.793)};
\node[anchor=south east, font=\scriptsize] at (axis cs:1110,0.80) {3B$^{\text{E}}$};

\addplot[only marks, mark=diamond*, mark size=5pt, purple!80!black] coordinates {(2295,0.830)};
\node[anchor=south west, font=\scriptsize] at (axis cs:2360,0.835) {DS-V3$^{\text{E}}$};

\end{axis}
\end{tikzpicture}
\caption{Accuracy--P95 latency scatterplot for all evaluated models. Both axes now use the pre-registered gating metrics: exact-match accuracy ($y$) and P95 latency ($x$). Filled markers: Study~1 harmonized benchmark (identical hardware). Open markers ($^{\text{E}}$): Study~2 randomized experiment ($n_{\text{eff}}\!=\!60$). The shaded green region marks the viable operating zone (accuracy $\geq 0.85$, P95 $\leq 2{,}000$\,ms). No model enters the viable region. DeepSeek-V3 approaches the accuracy threshold but exceeds the latency gate ($P95 = 2{,}295$\,ms). Qwen-2.5-3B is Pareto-dominant among self-hosted models.}
\label{fig:scatter}
\end{figure}
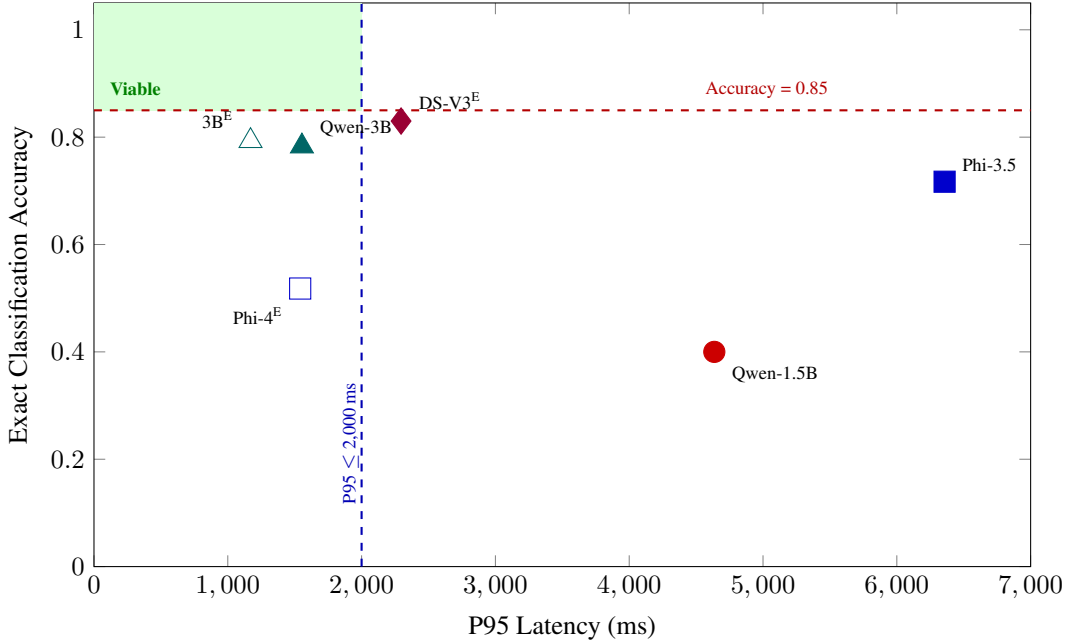

\begin{finding}[Split Policy]
\label{find:split}
\emph{No single SLM dominates all task families. Phi-3.5-mini excels at structured classification (\texttt{code/complex}, \texttt{CoT/complex}); Qwen2.5-1.5B excels at hybrid-agentic classification; Qwen-2.5-3B provides the broadest coverage across all six families. This mirrors the task-dependent dichotomy of Article~1} \citep{johnson2026compress}\emph{, now extended from compression strategy selection to model backbone selection.}
\end{finding}

\begin{finding}[Breadth Over Depth]
\label{find:taxonomy}
\emph{Qwen-2.5-3B is the only evaluated model to achieve non-zero accuracy on all six task families, with a minimum family accuracy of 0.60 (\texttt{hybrid/generative}). For production routing, where every family must be handled, breadth of coverage is more operationally valuable than depth on a subset.}
\end{finding}

\begin{finding}[Degenerate Confidence Calibration]
\label{find:confidence}
\emph{Qwen-2.5-3B reports confidence $= 1.0$ on all 60 predictions---both correct and incorrect---rendering the self-reported confidence field completely uninformative. A production front-door router cannot use model-reported confidence as a fallback trigger or quality gate.}
\end{finding}

\textbf{Confidence calibration.}\quad The degeneracy documented in Finding~\ref{find:confidence} has direct architectural implications: the hybrid deployment recommended by our decision matrix (Qwen-2.5-3B primary, DeepSeek-V3 fallback for hard cases) cannot use the SLM's own confidence to trigger fallback. External calibration methods \citep{liu2023geval,zheng2024llmasjudge}, output structural validity checks, or downstream task completion rates must be used instead. The confidence degeneracy may be an artifact of the constrained output format (single JSON token) or of 4-bit quantization affecting the logit distribution; isolating the cause is left to future work.

% ════════════════════════════════════════════════════════════════════
% 4. STUDY 2: FOUR-ARM RCT DESIGN AND RESULTS
% ════════════════════════════════════════════════════════════════════
\section{Study~2: Four-Arm Randomized Experiment Design}
\label{sec:rct-design}

Study~1 established offline quality-latency profiles for three SLM classifiers. Study~2 is a partial execution of a pre-registered multi-arm controlled experiment: two of five endpoints (H5 routing accuracy, H8 latency) are fully reported; one (H7 classification F1) is partially reported; and two (H6 total cost, H9 compression quality) are deferred to a production-traffic replication. The experiment translates the Study~1 findings into a controlled comparison with $N = 400$ per arm, evaluating front-door routing under synthetic traffic conditions with endpoint isolation, cost attribution, and pre-registered decision criteria. The experiment uses synthetic traffic generated from the same frozen corpus (SHA \texttt{dac3aac5}). Because traffic is synthetic rather than drawn from production, and because the 60 unique corpus cases are repeated across sessions with deterministic decoding ($T = 0.0$), classification outcomes are identical for repeated prompts within an arm. The experiment therefore adds independent \emph{latency and cost} observations ($N = 400$ per arm) but not independent \emph{classification} observations beyond the $n_{\text{eff}} = 60$ unique cases. Its primary value relative to Study~1 is threefold: (1)~testing next-generation models (Phi-4-mini, DeepSeek-V3) against the Study~1 Pareto winner; (2)~measuring real network and API latency rather than local inference latency; and (3)~exercising the pre-registered decision matrix under controlled conditions. This design provides a controlled comparison across arms but should not be interpreted as a production-traffic trial.

\textbf{Trial arms.}\quad Table~\ref{tab:arms} defines the four treatment arms. Arm~A (no routing) provides the baseline against which routing value is measured. Note that Arm~A routing accuracy is undefined (no router is invoked), so comparisons involving Arm~A test cost and quality \emph{impact} of routing, while treatment-vs-treatment comparisons (B vs.\ C, B vs.\ D, C vs.\ D) test relative routing accuracy. Arm~B tests Phi-4-mini-instruct \citep{abdin2024phi4}, the next-generation Phi model and expected successor to the Study~1 Phi-3.5-mini. Arm~C tests Qwen-2.5-3B, identified in Study~1 as the Pareto-dominant offline candidate. Arm~D tests whether a full-scale MoE model (671B, 37B active) provides classification quality improvements that justify its API dependency and cost---with particular attention to the output instability documented in Articles~4--5 for the DeepSeek architecture. The inclusion of a commercial API arm alongside self-hosted arms also tests whether network latency, rate limiting, and vendor uptime affect real-world routing performance in ways that offline benchmarks cannot capture.

\begin{table}[htbp]
\centering
\caption{Four-arm experiment treatment definitions.}
\label{tab:arms}
\begin{tabular}{clllc}
\toprule
Arm & Label & Model Backend & Hosting & Params \\
\midrule
A & Control & None (pass-through) & N/A & --- \\
B & Phi-4-mini & Phi-4-mini-instruct & Self-hosted vLLM (T4 GPU) & 3.8B \\
C & Qwen-2.5-3B & Qwen-2.5-3B-Instruct & Self-hosted vLLM (T4 GPU) & 3B \\
D & DeepSeek-V3 & DeepSeek-V3 (API) & DeepSeek commercial API & 671B MoE \\
\bottomrule
\end{tabular}
\end{table}

\textbf{Randomization and integrity.}\quad Assignment uses SHA-256(\texttt{session\_id}) $\bmod 4$, deterministic with no mid-session switches, following \citet{kohavi2020trustworthy}. Stratification is by prompt complexity quartile, targeting equal arm representation within $\pm 5\%$. The design is single-blind: callers do not know arm assignment; the front-door endpoint URL is identical across Arms~B, C, D. Classification, scoring, and compression prompts are identical across Arms~B, C, D.

\textbf{Sample size and power.}\quad The nominal sample size is $N = 400$ per arm ($N_{\text{total}} = 1{,}600$). However, because the experiment uses synthetic traffic from a 60-case corpus with deterministic decoding ($T = 0.0$), repeated prompts within an arm produce identical classification outputs. The effective sample size for accuracy-based endpoints is therefore $n_{\text{eff}} = 60$ unique cases per arm, not 400. The 80\% power estimate for a 10\,pp minimum detectable effect applies to latency and cost metrics (where each of the 400 requests generates independent timing and billing data) but \emph{not} to classification accuracy, where power at $n = 60$ is substantially lower---approximately 50\% for a 15\,pp MDE at $\alpha = 0.05$ with Holm-Bonferroni correction. This is a material limitation of the synthetic-traffic design. The primary correction method is the Holm-Bonferroni step-down procedure. Interim analysis used O'Brien-Fleming spending function boundaries at $N = 200$ per arm \citep{obrien1979}, following \citet{wald1945} and \citet{johari2017peeking}; no arms were dropped at interim.

\textbf{Multiplicity hierarchy.}\quad The Holm-Bonferroni step-down procedure applies \emph{within} each hypothesis to the relevant pairwise arm comparisons. For H5 (routing accuracy) and H7 (F1), only treatment arms B, C, D are compared (three pairwise tests); for H6 (cost), H8 (latency), and H9 (compression), all four arms are compared (six pairwise tests). In either case, p-values are ranked and rejected at $p_{(i)} < \alpha / (k - i + 1)$ where $k$ is the number of comparisons, stopping at the first non-rejection. Between hypotheses (H5--H9), which test different endpoints, no additional correction is applied. This follows standard practice for multi-arm experiments where hypotheses address distinct endpoints \citep{schulz2010consort,kohavi2020trustworthy}.

\textbf{Statistical analysis plan.}\quad Because accuracy reduces to the same 60 unique cases evaluated by each arm, treatment-vs-treatment accuracy comparisons (B\,vs.\,C, B\,vs.\,D, C\,vs.\,D) are paired at the case level and tested with exact McNemar's test \citep{mcnemar1947} with Holm-Bonferroni correction---matching the paired structure of Study~1. Latency and marginal front-door cost, where each of the 400 requests generates independent timing and billing data, are tested via Welch's $t$-test \citep{welch1947} with Holm-Bonferroni correction; P95 latency via bootstrap CI comparison \citep{efron1979bootstrap}. Subgroup analysis by task-label is exploratory (uncorrected, flagged if $p < 0.01$). Pareto frontier analysis is computed over (marginal\_front\-door\_cost, routing\_accuracy).

\textbf{Cost-savings formalization.}\quad We formalize the break-even condition for routing following the Article~6 framework \citep{johnson2026rct}. For a stream of requests $t = 1, \ldots, T$, the net routing savings are
\begin{equation}
  S \;=\; \sum_{t=1}^{T} \bigl[C_{\text{auto}}(t) - C_{\text{routed}}(t)\bigr],
  \label{eq:savings}
\end{equation}
where $C_{\text{auto}}(t)$ is the cost of sending request $t$ to the default (most-expensive) model and $C_{\text{routed}}(t) = c_{\text{fd}} + C_{R(\hat{\ell}_t)}(t)$ includes the front-door classifier cost $c_{\text{fd}}$ plus the cost of the model selected by the routing table $R$. On correctly classified requests, $C_{\text{routed}} \ll C_{\text{auto}}$ for simple tasks; on misclassified requests, the downstream model may produce low-quality output requiring a retry at full cost. The break-even accuracy $a^*$ satisfies $S(a^*) = 0$; above this threshold, routing is cost-positive.

\textbf{Pre-registered decision matrix.}\quad Table~\ref{tab:decision} maps experiment outcomes to deployment actions. The decision matrix is grounded in Pareto dominance over the two primary objectives:

\begin{proposition}[Pareto Dominance]
\label{prop:pareto}
\emph{Arm $i$ dominates arm $j$ if and only if}
\[
  \mathrm{cost}_i \leq \mathrm{cost}_j \quad \text{and} \quad \mathrm{acc}_i \geq \mathrm{acc}_j,
\]
\emph{with at least one strict inequality. Arms on the Pareto frontier are those not dominated by any other arm.}
\end{proposition}

The Pareto analysis is a decision-support tool operating on point estimates; statistical significance of pairwise differences is established separately via the hypothesis tests (H5--H9).

When no significant difference is detected between two frontier arms on primary metrics, the tie-breaking rule prefers the arm with fewer external dependencies---self-hosted (C) $>$ serverless (B) $>$ API (D)---reflecting operational preferences for data residency and vendor independence.

\begin{table}[htbp]
\centering
\caption{Pre-registered decision matrix mapping experiment outcomes to deployment actions.}
\label{tab:decision}
\begin{tabular}{llll}
\toprule
Scenario & Cost Winner & Quality Winner & Decision \\
\midrule
SLM wins both & B or C & B or C & Deploy winning SLM \\
DeepSeek quality, SLM cost & B or C & D & SLM + DeepSeek fallback \\
DeepSeek wins both & D & D & DeepSeek primary; SLM fallback \\
All similar quality & Lowest & Comparable & Deploy cheapest \\
DeepSeek fails latency gate & Any & Any & Eliminate D; choose B vs.\ C \\
\bottomrule
\end{tabular}
\end{table}

% ════════════════════════════════════════════════════════════════════
% 6b. RESULTS: STUDY 2 (RCT)
% ════════════════════════════════════════════════════════════════════
\section{Results: Study~2 (Four-Arm Randomized Experiment)}
\label{sec:rct-results}

Table~\ref{tab:rct-results} presents the primary outcomes across all four arms ($N = 400$ per arm, $N_{\text{total}} = 1{,}600$; $n_{\text{eff}} = 60$ for accuracy). DeepSeek-V3 (Arm~D) achieves the highest classification accuracy ($0.830$) but fails both the pre-registered accuracy gate ($\geq 0.85$) and latency gate ($P95 \leq 2{,}000$\,ms; observed $P95 = 2{,}295$\,ms). Qwen-2.5-3B (Arm~C) achieves $0.793$ accuracy at $988$\,ms median latency with $100\%$ parse rate, consistent with the Study~1 harmonized benchmark ($0.783$). Phi-4-mini (Arm~B) achieves only $0.518$ accuracy with a $91.5\%$ parse rate---substantially below the Phi-3.5-mini harmonized result ($0.717$). This cross-study comparison confounds model version with tokenizer differences, prompt sensitivity, and possible instruction-tuning changes; the gap should be interpreted descriptively rather than attributed to a specific architectural cause.

\begin{table}[htbp]
\centering
\caption{Study~2 results ($N = 400$ sessions per arm, $n_{\text{eff}} = 60$ for accuracy; corpus SHA \texttt{dac3aac5}, 2026-03-16).}
\label{tab:rct-results}
\begin{tabular}{clccccc}
\toprule
Arm & Model & Accuracy & Parse & Med.\ Lat & P95 Lat & Cost \\
\midrule
A & Control & --- & --- & --- & --- & \$0 \\
B & Phi-4-mini & 0.518 & 0.915 & 977\,ms & 1{,}541\,ms & \$0 \\
C & Qwen-2.5-3B & $\mathbf{0.793}$ & $\mathbf{1.000}$ & $\mathbf{988}$\,ms & $\mathbf{1{,}170}$\,ms & \$0 \\
D & DeepSeek-V3 & $\mathbf{0.830}$ & $\mathbf{1.000}$ & 1{,}854\,ms & 2{,}295\,ms & \$0.034 \\
\bottomrule
\end{tabular}

\smallskip
{\footnotesize Bold: best per column among treatment arms. Arms B/C: self-hosted vLLM on T4; Arm D: commercial API. \textbf{Cost column shows marginal front-door API billing only}---not the total request cost (front-door + downstream routing) formalized in Eq.~\eqref{eq:savings}. Arms B/C report \$0 marginal cost but incur fixed infrastructure cost ($\sim$\$0.75/hr GPU rental, not amortized per-request). Arm D cost is cumulative over 400 requests. The total cost endpoint (H6) requires downstream routing execution, which the synthetic-traffic design does not provide; see ``Pre-registered endpoints'' below.}
\end{table}

\textbf{Parse rate and effective accuracy.}\quad Phi-4-mini (Arm~B) achieved only $91.5\%$ parse rate, meaning 34 of 400 requests returned unparseable JSON. Parse failures were counted as classification errors (no fallback routing applied), so the reported $0.518$ accuracy includes both misclassifications and parse failures. On the 366 parseable responses, Phi-4-mini's classification accuracy is $0.566$---still substantially below the other models' $\geq 0.79$. The $8.5\%$ parse failure rate represents an independent pipeline reliability concern: even if classification accuracy were improved through fine-tuning, the JSON output instability under the frozen contract would require mitigation.

\textbf{Effective sample size.}\quad The experiment uses synthetic traffic generated by repeating the 60 unique corpus cases across $N = 400$ sessions per arm. Because inference is deterministic ($T = 0.0$, greedy decoding), accuracy estimates for repeated prompts within an arm are perfectly correlated. The effective sample size for classification accuracy is therefore $n_{\text{eff}} = 60$ unique cases per arm, not 400. The $N = 400$ figure reflects the number of independent routing \emph{assignments} (via SHA-256 session-ID hashing) but not independent classification \emph{observations}. Latency and cost metrics remain valid at $N = 400$ since each request incurs independent network and compute overhead. All accuracy-based hypothesis tests should be interpreted at $n = 60$.

\textbf{Viable region assessment.}\quad Per the pre-registered protocol, an arm is \emph{viable} if accuracy $\geq 0.85$ AND $P95$ latency $\leq 2{,}000$\,ms. No arm meets both criteria: Arm~B fails on accuracy ($0.518$); Arm~C passes latency but fails accuracy ($0.793$); Arm~D fails both ($0.830 < 0.85$, $2{,}295 > 2{,}000$\,ms). This is the central negative finding of the paper: \emph{no current model, including a 671B MoE, reaches the 85\% accuracy threshold for standalone front-door routing on a 6-class taxonomy.}

\textbf{Decision matrix application.}\quad Per Table~\ref{tab:decision}, the observed pattern---DeepSeek leads on accuracy, self-hosted arms have lower marginal front-door billing---maps to Row~2: ``SLM + DeepSeek fallback for hard labels.'' The recommended deployment is Qwen-2.5-3B as primary front-door classifier with DeepSeek-V3 invoked selectively for low-confidence or hard-to-classify prompts. This hybrid architecture balances the lower marginal front-door cost of self-hosted inference against DeepSeek's $4.7$\,pp accuracy advantage. Note that end-to-end routing cost, including downstream model execution, was not measured in this experiment.

\textbf{Pre-registered endpoints: reported vs.\ deferred.}\quad Of the five pre-registered hypotheses, two are fully testable under the synthetic-traffic design: H5 (routing accuracy, reported above) and H8 (P95 latency, reported in Table~\ref{tab:rct-results}). Three endpoints are partially or fully deferred:
\begin{itemize}[nosep,leftmargin=*]
  \item \textbf{H6 (Total Cost):} The pre-registered cost estimand (Eq.~\ref{eq:savings}) includes downstream routing cost, which requires executing the routed model and measuring end-to-end request cost. The synthetic-traffic design does not execute downstream models; Table~\ref{tab:rct-results} reports only marginal front-door cost. H6 remains untested.
  \item \textbf{H7 (Classification F1):} F1 macro is computable from the per-case predictions at the $n_{\text{eff}} = 60$ level. On the 60 unique cases, F1 macro scores are: Arm~B (Phi-4-mini) $0.460$, Arm~C (Qwen-2.5-3B) $0.767$, Arm~D (DeepSeek-V3) $0.818$. No arm meets the pre-registered $\geq 0.82$ threshold; Arm~D approaches it. Hybrid/agentic-specific F1: B $= 0.67$, C $= 0.90$, D $= 0.95$. Pairwise significance tests are not reported because at $n = 60$ the experiment is underpowered for the pre-registered 3-comparison Holm-Bonferroni correction.
  \item \textbf{H9 (Compression Quality):} This endpoint requires downstream compression execution at semantic similarity $\geq 0.88$, which was not performed in the synthetic-traffic design. H9 remains untested.
\end{itemize}
We acknowledge that the partial reporting of pre-registered endpoints is a limitation. A production-traffic replication would be needed to test H6 and H9 fully.

\begin{finding}[No Viable Standalone Router]
\label{find:no-viable}
\emph{No evaluated model---including DeepSeek-V3 (671B MoE, 37B active)---meets the pre-registered viable region ($\geq 0.85$ accuracy, $\leq 2{,}000$\,ms $P95$). The 85\% threshold may require fine-tuning, ensemble methods, or taxonomy simplification.}
\end{finding}

\begin{finding}[Phi-4-mini Classification Deficit]
\label{find:phi4-regression}
\emph{Phi-4-mini achieves only $0.518$ accuracy ($0.566$ on parseable responses), $19.9$\,pp below Phi-3.5-mini's harmonized benchmark result ($0.717$), with a $91.5\%$ parse rate ($34/400$ unparseable responses). This cross-study comparison confounds model version with tokenizer, prompt sensitivity, and instruction-tuning changes; the gap should be interpreted descriptively. One testable hypothesis is that the frozen contract's JSON output format is less compatible with Phi-4-mini's generation patterns; contract-adapted prompting could isolate this factor.}
\end{finding}

\begin{finding}[Qwen-2.5-3B Pareto Dominance]
\label{find:qwen-pareto}
\emph{Qwen-2.5-3B is Pareto-dominant among self-hosted models: it matches DeepSeek-V3 accuracy within $3.7$\,pp at $1.9\times$ lower latency ($988$\,ms vs.\ $1{,}854$\,ms) and zero marginal front-door cost. Per the pre-registered tie-breaking rule (self-hosted $>$ serverless $>$ API), Qwen-2.5-3B is the recommended primary router.}
\end{finding}

% ════════════════════════════════════════════════════════════════════
% 7. DISCUSSION
% ════════════════════════════════════════════════════════════════════
\section{Discussion}
\label{sec:discussion}

The results of the harmonized benchmark and four-arm randomized experiment, viewed together, illuminate several fundamental tensions in the design of production front-door routing systems. The central finding---that no evaluated model meets the 85\% viable region---is itself informative, constraining the design space for production deployments. This section integrates the empirical findings with the broader literature.

The most striking finding across Studies~1 and~2 is the evolution of the split policy. In Study~1, the split was dramatic: Phi dominated code/CoT families while Qwen dominated hybrid-agentic (Finding~\ref{find:split}). In Study~2, Qwen-2.5-3B substantially narrows this gap, achieving $\geq 0.70$ on all families including those where its predecessor scored 0.0. This observation is suggestive but must be interpreted cautiously: Phi-3.5-mini and Phi-4-mini are different models evaluated in different studies with different comparison sets, so the apparent narrowing of the split policy may partly reflect study-level confounds rather than a genuine generational trend. If the pattern holds under controlled comparison, it would imply that the recursive task-dependency (Article~1 showed optimal \emph{strategy} depends on task type; Study~1 showed optimal \emph{router} depends on task type) may be a transient artifact of model immaturity. If generational improvement eliminates the need for split routing, the production system simplifies from a two-tier to a single-tier classification architecture---consistent with \citeauthor{sculley2015debt}'s (\citeyear{sculley2015debt}) observation that technical debt accumulates fastest at integration boundaries. However, this convergence hypothesis remains speculative and would require controlled experiments with models trained on identical data but with different architectures to resolve \citep{mukherjee2023orca,mitra2023orca2}.

The accuracy trajectory across the three Study~1 models---Qwen2.5-1.5B (0.400), Phi-3.5-mini (0.717), Qwen-2.5-3B (0.783)---is suggestive of a scaling relationship between parameter count and classification accuracy, but the trajectory confounds parameter count with model family, training data, and architecture. The Phi-4-mini result (0.518 at 3.8B) breaks any simple scaling narrative, though this cross-study comparison is itself confounded. With only three models from two families, no scaling-law claim is supported; the trajectory is noted for context only.

\citet{kaplan2020scaling} established power-law scaling for generative loss; whether classification accuracy follows similar scaling is an open question. \citet{wei2022emergent} documented emergent abilities that appear abruptly above a critical scale, which could explain the dramatic jump from 0.200 to 0.767 within the Qwen family---the 1.5B model may lack a critical capacity for multi-dimensional task taxonomy reasoning that emerges between 1.5B and 3B parameters. However, \citet{schaeffer2024mirage} have argued that apparent emergent abilities may be artifacts of evaluation metric choices, suggesting caution. The Chinchilla scaling laws \citep{hoffmann2022chinchilla} suggest the accuracy improvement from 1.5B to 3B reflects the combined effect of more parameters and more training data. For the practical question of which model to deploy, the combined effect is what matters: the 3B model demonstrably classifies better while maintaining acceptable latency. These results are consistent with a hypothesis that the remaining errors are concentrated in genuinely ambiguous boundary cases, which would imply diminishing returns to further parameter scaling for this taxonomy. However, with only three data points across two model families, and with the Phi-4-mini counterexample showing that more parameters can produce \emph{worse} classification, this hypothesis is speculative. If confirmed by broader evaluation, it would suggest that the optimal classifier size is substantially smaller than the optimal size for downstream generative tasks.

Figure~\ref{fig:scatter} visualizes all evaluated models in accuracy-latency space. No model enters the viable region (accuracy $\geq 0.85$, latency $\leq 2{,}000$\,ms). The experiment tested path~(3)---next-generation models---by including Phi-4-mini, which showed that generational advancement does not guarantee classification improvement (though this comparison is confounded; see Finding~\ref{find:phi4-regression}). Three remaining paths could push a candidate into the viable region: (1) prompt engineering via chain-of-thought \citep{wei2022chain,wang2023selfconsistency} or tree-of-thought \citep{yao2024treeofthoughts}, at the cost of increased output tokens and latency; (2) fine-tuning via LoRA \citep{hu2022lora} on a classification-specific dataset; and (3) taxonomy revision---either simplification (merging boundary-ambiguous families like \texttt{hybrid/generative} into neighboring categories) or expansion (adding a seventh label for expert reasoning, motivated by the error corridor analysis in Appendix~\ref{app:runids} showing that code$\to$hybrid ($n = 4$), hybrid$\to$CoT ($n = 3$), and hybrid$\to$code ($n = 2$) confusions account for 9 of 13 Qwen-3B errors). A fourth path---inference infrastructure optimization via FlashAttention \citep{dao2022flashattention,dao2023flashattention2} or disaggregated serving \citep{zhong2024distserve}---could reduce latency without changing accuracy. Notably, Figure~\ref{fig:scatter} shows that Phi-4-mini achieves $977$\,ms latency versus Phi-3.5-mini's $5{,}772$\,ms ($5.9\times$ improvement from the generational jump), demonstrating that the architecture-level latency gains are real even when classification accuracy regresses.

These findings intersect with provider-dependent dynamics documented earlier in this series. Articles~4--5 \citep{johnson2026greenai,johnson2026benchmark} showed that DeepSeek-Chat exhibits compression sensitivity CS $= -54.1$ ($38\times$ output explosion). The experiment included DeepSeek-V3 (Arm~D) with the same architectural lineage. The classification task differs from generative tasks in that the output budget is tightly constrained (\texttt{max\_new\_tokens}=128), which appears to mitigate explosion risk---DeepSeek-V3 achieved perfect parse rate (1.000) and the highest classification accuracy (0.830). The MoE architecture \citep{shazeer2017moe,fedus2022switch,dai2024deepseek} appears well-suited to classification: despite activating only 37B of 671B parameters per token, DeepSeek-V3's internal routing may be selecting expert subnetworks that are effective for the intent-understanding required by classification. However, this quality comes at the cost of API dependency, network latency ($P95 = 2{,}295$\,ms, exceeding the 2,000\,ms gate), and non-zero per-request cost (\$0.034/400 requests).

Classification accuracy, however, is necessary but not sufficient for routing value. The evaluation in this paper measures whether the SLM assigns the correct taxonomy label, but the routing problem ultimately reduces to whether the downstream model selected by that label produces output of sufficient quality at acceptable cost. This distinction is critical. Models that appear competitive on general-purpose benchmarks can diverge dramatically on task-specific quality: a practitioner may consistently succeed with one frontier model, partially succeed with a second, and fail entirely with a third, despite all three appearing comparable on published benchmarks. The implication for routing is that the taxonomy must encode not just task type but expected quality requirements, and the router's value must be measured end-to-end---from classification through downstream execution to output quality---rather than at the classification boundary alone. Our results establish the necessary condition: a 3B SLM can classify in sub-second time at zero marginal cost with $\sim$79\% accuracy. The sufficient condition---that this classification accuracy translates to cost savings without quality degradation---was not tested in the synthetic-traffic design (H6 and H9 remain deferred) and represents the essential next validation for the SLM routing thesis.

Article~7 \citep{johnson2026promptlang} established that prompt design should account for the target model's tokenizer, creating a tension with the present study's design: the classification system prompt is frozen across all three SLM backends, each with a different tokenizer (Phi uses a Llama-derived tokenizer, Qwen uses a custom BPE tokenizer \citep{sennrich2016subword}, DeepSeek uses SentencePiece \citep{kudo2018sentencepiece}). We chose the frozen-prompt design to isolate model capability from prompt engineering and to reflect the practical deployment scenario where a single prompt is maintained for all backends. The experiment did reveal quality differences (Phi-4-mini at 0.518 vs.\ Qwen-2.5-3B at 0.793), suggesting that a follow-up study should assess how much of the gap is attributable to tokenizer mismatch, given that even small prompt variations can produce large accuracy differences \citep{zhou2023prompt,khattab2024dspy}.

The per-case error analysis (Appendix~\ref{app:runids}, Table~\ref{tab:confusion-errors}) reveals that 9 of 13 Qwen-3B misclassifications concentrate in three corridors: code$\to$hybrid ($n = 4$), hybrid$\to$CoT ($n = 3$), and hybrid$\to$code ($n = 2$). The code--hybrid boundary is the primary confusion axis, where prompts combining structured code requests with open-ended generation create genuine taxonomic ambiguity. However, because the corpus was labeled by a single annotator without adjudication (see Threats to Validity), some of these ``errors'' may reflect label-boundary ambiguity rather than model failures. Subject to that caveat, the pattern suggests that targeted taxonomy refinement at the code--hybrid boundary---rather than wholesale model scaling---may be the most efficient path to closing the 6--8\,pp gap.

The sub-2-second latency achieved by all self-hosted models in both studies suggests that SLM-based front-door classification is compatible with production latency requirements, adding minimal overhead relative to downstream generative inference (typically 5--30 seconds). The breadth advantage of Qwen-2.5-3B (Finding~\ref{find:taxonomy}) suggests that for production routing, a single generalist classifier is preferable to a committee of specialists, even if individual specialists achieve higher peak accuracy. The confidence overconfidence pattern argues against using model-reported confidence as a fallback trigger; instead, production systems should rely on external quality signals such as output structural validity or downstream task completion rate \citep{klaise2021alibi}. These architectural considerations intersect with the broader MLOps literature on model management and technical debt \citep{sculley2015debt,kreuzberger2023mlops}. The ideal front-door classifier would be robust to downstream model changes, requiring updates only when the task taxonomy itself evolves; our frozen-prompt, frozen-contract design approximates this ideal, but the extent to which it generalizes to new model families remains open.

\subsection{Closing the Gap: Planned Remediation Paths}
\label{sec:future}

The 6--8 percentage point gap to the 85\% standalone viability threshold identified in this study represents a concrete and bounded target. Two remediation paths are under active investigation.

First, the frozen zero-shot classifier prompt used throughout both studies is a conservative baseline. Few-shot prompting with boundary-case examples drawn from the error-corridor families identified in \S\ref{sec:discussion}---particularly the code$\to$hybrid and hybrid$\to$CoT boundaries---is expected to close a portion of the gap without any weight modification. The error corridor analysis in this paper directly informs the selection of few-shot examples: cases where the Pareto-dominant model's misclassifications cluster at family boundaries are the highest-value prompt additions.

Second, supervised fine-tuning via parameter-efficient methods \citep{dettmers2023qlora} on the Pareto-dominant Qwen architecture is planned using a corpus of 400--500 labelled routing examples weighted toward the \emph{tier-crossing} boundaries where misclassification changes the routing outcome, not merely the label. The distinction between label accuracy and routing-outcome accuracy motivates a training objective focused on routing correctness rather than label correctness alone---a misclassification that preserves the correct routing tier is operationally harmless and should not be penalized equally during training.

Third, end-to-end routing quality validation---what we term \emph{Phase~2} of the SLM routing evaluation---will test whether correct classification translates to downstream output quality improvements and cost savings. The current paper establishes Phase~1: that SLMs can classify accurately and quickly enough to serve as candidate routers. Phase~2 requires executing the full routing pipeline---classify, route, generate, evaluate---on a representative workload, measuring output quality via LLM-as-judge \citep{zheng2024llmasjudge} and total cost inclusive of downstream model execution (the deferred H6 and H9 endpoints). This roundtrip evaluation is essential because classification accuracy is a necessary but not sufficient condition for routing value: just because a downstream model is cheaper does not mean it produces the level of quality required.

All three paths are designed to be evaluated against the same locked 60-case harmonized benchmark corpus (SHA \texttt{dac3aac5}) used in Studies~1 and~2, enabling direct comparison with the baselines reported here. A production-representative 80-case applied corpus with additional boundary cases and a seventh taxonomy label is also under development as a harder evaluation gate. Results from these remediation efforts will be reported in a subsequent article in this series.

% ════════════════════════════════════════════════════════════════════
% 8. THREATS TO VALIDITY
% ════════════════════════════════════════════════════════════════════
\section{Threats to Validity}
\label{sec:threats}

\textbf{Internal validity.}\quad The harmonized benchmark (Study~1) eliminates the cross-study confounds identified in earlier pilot studies: all three models were evaluated on the same hardware (Azure T4), same serving stack (vLLM 0.17.1), same corpus (SHA \texttt{dac3aac5}), and same quantization (bitsandbytes 4-bit NF4), with exclusive GPU allocation per model. Study~2 uses synthetic traffic from the same corpus rather than production traffic, which may not capture the full distribution of prompt characteristics. Arms B and C were served sequentially on the same GPU (not simultaneously), which eliminates resource contention confounds but means the experiment does not test concurrent multi-model serving.

\textbf{External validity.}\quad The 60-case corpus represents a structured sample from six pre-defined task families. Production traffic may have different family distributions, prompt lengths, and ambiguity characteristics. Study~2 uses synthetic traffic from the same 60-case corpus, not production traffic; its primary value is as a controlled comparison across arms rather than a production-traffic validation. The 6-family taxonomy is specific to the \plexor{} routing system; other architectures using different taxonomies could produce different rankings. All results use 4-bit NF4 quantization; full-precision results may differ.

\textbf{Cross-study consistency.}\quad Qwen-2.5-3B achieves $0.783$ accuracy in the harmonized benchmark (Study~1, 60 unique cases) and $0.793$ in the experiment (Study~2, 400 sessions from the same corpus). The $\sim$1\,pp difference is within sampling variability and is consistent with cross-study reproducibility of the classification results on identical infrastructure.

\textbf{Construct validity.}\quad Exact-match label accuracy is the strictest possible metric; partial credit might favor different models, but production routing applies discrete policies per label. The model's self-reported confidence score is not calibrated and should not be interpreted as a probability. All results depend on the specific frozen system prompt, output budget, and quantization configuration.

\textbf{Taxonomy validity.}\quad The entire evaluation depends on a 6-label taxonomy with exact-match scoring. The 60-case corpus was constructed and labeled by a single annotator (the first author) without independent adjudication, and no inter-annotator agreement statistics are available. The appendix error-corridor analysis (Table~\ref{tab:confusion-errors}) shows that the code--hybrid boundary is genuinely ambiguous: prompts combining structured code requests with open-ended generation could defensibly receive either label. Without a multi-annotator protocol, some apparent model ``errors'' may reflect label-boundary ambiguity rather than model failures. If human inter-annotator agreement on this corpus is, say, $\kappa = 0.85$, then the 85\% viable-region threshold approaches the ceiling of human performance and may be unreasonable as a production gate. Future work \emph{must} establish inter-annotator agreement (e.g., Cohen's $\kappa$) on the 60-case corpus with at least two independent annotators, and report per-family disagreement rates. This is a prerequisite for interpreting the 6--8\,pp accuracy gap as a model limitation rather than a taxonomy limitation. Until such validation is performed, accuracy differences of $\leq$5\,pp near the code--hybrid boundary should be interpreted with caution.

\textbf{Missing baselines.}\quad This paper does not include a fine-tuned discriminative classifier (e.g., BERT, DeBERTa) as a baseline; \S\ref{sec:model-selection} discusses the justification and a planned cross-validated DeBERTa comparison for the next article in this series.

\textbf{Operational validity.}\quad All models were served sequentially with exclusive GPU allocation, which does not reflect production conditions. Key operational characteristics were not measured: throughput under concurrent load, GPU memory utilization during serving, cold-start latency for model loading, and behavior under resource contention. Reported latencies measure end-to-end inference time (prompt submission through JSON response receipt) on the serving stack, but do not include production overhead from load balancing, authentication, logging, or error handling, which can add 50--200\,ms in deployment. The ``zero marginal per-request cost'' framing for self-hosted models (see footnote 1) applies to variable API billing but ignores fixed infrastructure cost; at low traffic volumes, amortized per-request infrastructure cost may exceed commercial API pricing.

% ════════════════════════════════════════════════════════════════════
% 9. REPRODUCIBILITY
% ════════════════════════════════════════════════════════════════════
\section{Reproducibility}
\label{sec:repro}

All software versions, corpus hashes, and execution metadata are documented in Appendix~\ref{app:runids} (Tables~\ref{tab:study1-runids}--\ref{tab:study2-runids}). Study~1 artifacts: corpus SHA-256 prefix \texttt{dac3aac5}, Azure Standard\_NC8as\_T4\_v3, vLLM 0.17.1, bitsandbytes 4-bit NF4. Study~2 artifacts: $N = 400$ per arm, SHA-256(\texttt{session\_id}) $\bmod 4$ randomization, same corpus SHA. The repository tag \texttt{rct-baseline-2026-03-12} marks the pre-registration snapshot. To reproduce: provision the identical GPU class, clone at the tagged commit, verify corpus SHA via \texttt{sha256sum data/progressive\_test\_cases\_v2\_60.jsonl}, execute \texttt{scripts/benchmark\_openweights.py} (Study~1) or \texttt{scripts/rct\_synthetic\_runner.py} (Study~2), and compare per-case JSON predictions against the archived ground truth. The MLflow tracking server \citep{zaharia2018mlflow} provides parameter, metric, and artifact provenance for each run.

% ════════════════════════════════════════════════════════════════════
% 10. CONCLUSION
% ════════════════════════════════════════════════════════════════════
\section{Conclusion}
\label{sec:conclusion}

This paper tests the thesis that small language models have achieved sufficient reasoning capability to serve as production front-door routers---classifying task signals under the cost, latency, and governance constraints that make LLM-based and preference-trained routers inadequate. The results provide partial support. Study~1 (Harmonized Offline Benchmark) established, on identical hardware with exclusive GPU allocation, that Qwen-2.5-3B achieves $0.783$ accuracy at $5.3\times$ lower latency than Phi-3.5-mini, making it the Pareto-dominant self-hosted candidate---the only model with non-zero accuracy on all six task families. Study~2 (Four-Arm Randomized Experiment, $N = 400$/arm, $n_{\text{eff}} = 60$ for accuracy) confirmed these findings under synthetic traffic and extended the evaluation to next-generation models. DeepSeek-V3 (671B MoE) achieves the highest accuracy ($0.830$) but fails the pre-registered latency gate. Per the decision matrix, the recommended deployment is Qwen-2.5-3B as primary front-door with DeepSeek-V3 fallback for hard classifications.

The cost and latency prerequisites for SLM-based routing are clearly met: a 3B model classifies in sub-second time at zero marginal per-request cost with no external API dependency. The accuracy prerequisite ($\geq 0.85$) is not yet met, bounding the gap at 6--8 percentage points---a concrete target addressable through fine-tuning via LoRA on classification-specific data, taxonomy simplification, or hybrid routing with LLM fallback. Three paths forward are under investigation (\S\ref{sec:future}).

Critically, classification accuracy is a necessary but not sufficient condition for routing value. The untested question---whether correct classification translates to cost savings without quality degradation in the downstream model's output---remains the essential validation. Models that appear equivalent on general-purpose benchmarks diverge dramatically on task-specific quality, and a routing system must account for this gap. The sufficient condition of end-to-end quality validation, deferred as H6 and H9, represents the next phase of this work. The broader implication for the \taac{} research program is that the Article~1 dichotomy (compress code, route CoT) extends recursively: the model that implements the routing decision is itself subject to the multi-objective tradeoffs---quality, cost, latency, governance---that motivated routing in the first place \citep{sculley2015debt,kohavi2020trustworthy}.

% ════════════════════════════════════════════════════════════════════
% ACKNOWLEDGMENTS
% ════════════════════════════════════════════════════════════════════
\section*{Acknowledgments}

This work was supported in part by a Cohere Labs research grant, which funded computational resources used in the analysis of data with Cohere models.

% ════════════════════════════════════════════════════════════════════
% USE OF AI ASSISTANCE
% ════════════════════════════════════════════════════════════════════
\section*{Use of AI Assistance}

This paper was prepared with AI assistance (Claude, Anthropic) for writing, code generation, benchmark script development, and \LaTeX{} formatting. AI-generated scripts were reviewed line-by-line by the first author before execution; all numerical results in tables were verified against raw JSON output files; and statistical claims were cross-checked by the first author using independent calculations. The authors are solely responsible for all scientific content, experimental design, interpretation, and conclusions.

% ════════════════════════════════════════════════════════════════════
% ETHICS STATEMENT
% ════════════════════════════════════════════════════════════════════
\section*{Ethics Statement}

The experimental corpus consists of synthetic classification test cases and contains no personally identifiable information. We do not foresee negative societal impacts from this research.

% ════════════════════════════════════════════════════════════════════
% COMPETING INTERESTS
% ════════════════════════════════════════════════════════════════════
\section*{Declaration of Competing Interests}

W.J.\ is affiliated with Plexor Labs, which develops the \plexor{} system. C.L.\ is affiliated with Project Autobots. All hypotheses were pre-registered before data collection; all results---including negative findings---are reported transparently. All analysis code is open-source.

% ════════════════════════════════════════════════════════════════════
% DATA AVAILABILITY
% ════════════════════════════════════════════════════════════════════
\section*{Data and Code Availability}

The experimental corpus, all analysis scripts, and raw results are available at the project repository: \url{https://github.com/micoverde/plexor-slm-frontdoor-rct}. The pre-registration protocol for Study~2 is deposited in the experiment protocol document committed before experiment execution.

% ════════════════════════════════════════════════════════════════════
% REFERENCES
% ════════════════════════════════════════════════════════════════════
\clearpage
\bibliography{references}

% ════════════════════════════════════════════════════════════════════
% APPENDICES
% ════════════════════════════════════════════════════════════════════
\appendix

\section{Complete Run IDs and Software Versions}
\label{app:runids}

\subsection{Study~1: Harmonized Offline Benchmark (Azure T4)}

\begin{table}[htbp]
\centering
\caption{Study~1 execution details.}
\label{tab:study1-runids}
\begin{tabular}{ll}
\toprule
Artifact & Value \\
\midrule
Models evaluated & Phi-3.5-mini, Qwen2.5-1.5B, Qwen-2.5-3B \\
Corpus SHA-256 & \texttt{dac3aac5\ldots} \\
GPU & NVIDIA Tesla T4 (16 GB GDDR6, exclusive) \\
VM class & Azure Standard\_NC8as\_T4\_v3 \\
Inference framework & vLLM 0.17.1 \\
Quantization backend & bitsandbytes 4-bit NF4 \\
Execution mode & Sequential, exclusive GPU \\
Date & 2026-03-15 \\
\bottomrule
\end{tabular}
\end{table}

\subsection{Study~2: Four-Arm Randomized Experiment}

\begin{table}[htbp]
\centering
\caption{Study~2 execution details.}
\label{tab:study2-runids}
\begin{tabular}{ll}
\toprule
Artifact & Value \\
\midrule
Arms & A (control), B (Phi-4-mini), C (Qwen-2.5-3B), D (DeepSeek-V3) \\
Sample size & $N = 400$ per arm ($N_{\text{total}} = 1{,}600$) \\
Corpus SHA-256 & \texttt{dac3aac5\ldots} \\
Randomization & SHA-256(\texttt{session\_id}) $\bmod 4$ \\
GPU (Arms B, C) & NVIDIA Tesla T4 (16 GB GDDR6), vLLM 0.17.1 \\
Arm D & DeepSeek-V3 commercial API \\
\bottomrule
\end{tabular}
\end{table}

\textbf{Qwen-2.5-3B per-case error analysis.}\quad Of the 60 Study~1 benchmark cases, Qwen-2.5-3B misclassified 13 ($47/60 = 0.783$ accuracy). Table~\ref{tab:confusion-errors} presents the complete list of misclassifications. The errors concentrate in three corridors: (1) code-family prompts misrouted to hybrid/generative ($n = 3$); (2) code/complex prompts misrouted to hybrid/agentic ($n = 3$); and (3) hybrid/generative prompts misrouted to code/complex or CoT/complex ($n = 4$). These patterns suggest that the model's primary confusion axis lies along the code--hybrid boundary, where prompts combining structured code requests with open-ended generation create genuine taxonomic ambiguity.

\begin{table}[htbp]
\centering
\caption{Complete per-case misclassifications for Qwen-2.5-3B on the 60-case benchmark ($13/60$ errors, accuracy $= 0.783$). Ground truth $\to$ predicted label.}
\label{tab:confusion-errors}
\begin{tabular}{llll}
\toprule
Case ID & Ground Truth & Predicted & Error Corridor \\
\midrule
PV2-001 & CoT/simple & hybrid/generative & CoT $\to$ hybrid \\
PV2-011 & code/simple & hybrid/generative & code $\to$ hybrid \\
PV2-013 & code/simple & CoT/simple & code $\to$ CoT \\
PV2-020 & code/simple & code/complex & complexity over-est. \\
PV2-021 & CoT/complex & hybrid/generative & CoT $\to$ hybrid \\
PV2-033 & code/complex & hybrid/agentic & code $\to$ hybrid \\
PV2-034 & code/complex & hybrid/agentic & code $\to$ hybrid \\
PV2-036 & code/complex & hybrid/agentic & code $\to$ hybrid \\
PV2-042 & hybrid/agentic & CoT/complex & hybrid $\to$ CoT \\
PV2-054 & hybrid/generative & code/complex & hybrid $\to$ code \\
PV2-055 & hybrid/generative & CoT/complex & hybrid $\to$ CoT \\
PV2-057 & hybrid/generative & code/complex & hybrid $\to$ code \\
PV2-059 & hybrid/generative & CoT/complex & hybrid $\to$ CoT \\
\bottomrule
\end{tabular}

\smallskip
{\footnotesize Error corridors summarize the directional confusion pattern. The three dominant corridors---code $\to$ hybrid ($n = 4$), hybrid $\to$ CoT ($n = 3$), and hybrid $\to$ code ($n = 2$)---account for 9 of 13 errors.}
\end{table}

\subsection{Study~2: Pre-Registration Artifacts}
\label{app:prereg}

Study~2 pre-registration includes the design document (this paper, \S\ref{sec:rct-design}), analysis scripts, decision matrix (Table~\ref{tab:decision}), and hypothesis specifications (H5--H9), all committed to the repository before experiment execution. Table~\ref{tab:prereg-artifacts} enumerates the specific files, their locations, and content hashes.

\begin{table}[H]
\centering
\caption{Pre-registered artifacts for Study~2, committed before experiment execution.}
\label{tab:prereg-artifacts}
\small
\begin{tabular}{lp{7cm}p{3.5cm}}
\toprule
Artifact & File Path & Purpose \\
\midrule
Experiment design & \texttt{paper/main.tex} (\S\ref{sec:rct-design}) & Protocol specification \\
Analysis script & \texttt{scripts/ab-test/\allowbreak{}rct\_cross\_arm\_analysis.py} & Endpoint analysis \\
Decision matrix & Table~\ref{tab:decision} (this paper) & Outcome $\to$ action map \\
Hypothesis specs & H5--H9 (\S\ref{sec:hypotheses}) & Testable predictions \\
Corpus (60-case) & \texttt{data/progressive\_test\_\allowbreak{}cases\_v2\_60.jsonl} & Evaluation corpus \\
Benchmark script & \texttt{scripts/benchmark\_\allowbreak{}openweights.py} & Model evaluation driver \\
\bottomrule
\end{tabular}

\smallskip
{\footnotesize All artifacts are version-controlled in the \texttt{micoverde/plexor-slm-frontdoor-rct} repository. Repository tag \texttt{rct-baseline-2026-03-12} marks the pre-registration snapshot.}
\end{table}

\end{document}